\documentclass[12pt]{iopart}
\usepackage{iopams}
\usepackage{epsf}

\begin{document}

\title{A fractal set from the binary reflected Gray code}

\author{J A Oteo\dag\ and J.
Ros\ddag\ \footnote[3]{To whom correspondence should be addressed} }

\address{\dag\ Departament de F\'{\i}sica Te\`{o}rica, Universitat de
Val\`{e}ncia,  46100-Burjassot, Val\`{e}ncia, Spain}

\address{\ddag Departament de F\'{\i}sica Te\`{o}rica  and Instituto de
F\'{\i}sica Corpuscular, Universitat de Val\`{e}ncia,
46100-Burjassot, Val\`{e}ncia, Spain}

\ead{oteo@uv.es, rosj@uv.es}

\begin{abstract}
{The permutation associated with the decimal expression of the
binary reflected Gray code with $N$ bits is considered. Its cycle
structure is studied. Considered as a set of points, its
self-similarity is pointed out. As a fractal, it is shown to be the
attractor of a IFS. For large values of $N$ the set is examined from
the point of view of  time series analysis.}
\end{abstract}



%\submitto{\JPA }



\maketitle%\documentclass[ukenglish,a4paper]{article}

\section{Introduction}
In the most general setting one can define a Gray code as a listing
of \emph{words}, formed with $N$ \emph{letters} from an
\emph{alphabet} with $k$ \emph{symbols}, ordered in such a way that
any two successive entries differ in just one symbol. These codes
are named after Frank Gray, an engineer at Bell Laboratories who
employed it in a patent in 1953 \cite{ma}, although the use of the
code has been traced back at least to 1878 when the French engineer
Jean Maurice Emile Baudot (after whom the name baud for the
signaling rate measure was chosen) developed a telegraph code on
these grounds which readily prevailed in France.

We are interested in the case where numerical symbols are involved.
We can still distinguish different types of codes according to the
number system used. The most commonly employed basis are of course 2
and 10 giving rise to the \emph{binary} and \emph{decimal} Gray
codes respectively. Originally, the characteristic feature of these
codes was important in mechanical devices such as counters, for no
more than one piece at a time needs to act in every counting
operation. Needless to say, this feature decreases the possibility
of occurrence of errors albeit these may still be large. Although
nowadays perhaps old-fashioned, a very illustrative and familiar
situation comes from the mechanical odometer of a car. When it reads
9999 km. five inner wheels have to rotate to show 10000 km. Gray
code is designed to avoid such {\sl multiple} changes.

In a wide variety of types Gray codes have been and still are
applied in a large range of extremely distinct fields. We leave
apart their use as puzzle-solvers \cite{Hanoi} (e.g. the Towers of
Hanoi or the Chinese rings) or in some even more bizarre application
like campanology \cite{campana}. As a sample of more canonical
fields where Gray codes are relevant we can cite as mere examples:
database searching \cite{database}, design of charged particle
detectors \cite{hodoscopio}, space filling Hilbert curves
\cite{hilbert}, genetic algorithms \cite{ga} and their use in
parameter optimization problems \cite{Rowe}, neural networks
\cite{neural}, computation of unstable periodic orbits in chaotic
attractors \cite{orbits} and, very recently, quantum computing
\cite{Finland}. Even a little surfing in the web will convince the
reader of their importance in commercial matters related to A-D
converters or design of communication codes.

In our case it was using the native Gray code to generate Walsh
functions that we were led to the considerations presented in this
paper. We take a Gray code as the play ground in which we build and
represent some construction that is then analyzed by techniques
familiar in other fields like dynamical systems or time series
analysis but whose application is novel, so we think, in this
context. The construction we refer to is a permutation, $P_G, $ of
natural numbers which in fact constitutes a known sequence, listed
as A003188 in OEIS \cite{sloane}. Our approach is basically
descriptive and is not directly related to the interesting topic  of
the mutual relationship between Gray codes and permutations which
has been extensively considered in the literature \cite{art1}. In
Section \ref{Reflec} the specific type of Gray code which we will
focus our attention on is briefly reviewed, namely the so-called
\emph{binary reflected Gray code} (BRGC for short). Then we build
what we call Gray curve which we analyze from a fractal geometry
point of view in Section \ref{selfsim}. Statistical considerations
borrowed from modern time series analysis are applied in Section
\ref{timeseries}. For easier reading we have postponed to Section
\ref{permu} the discussion of the technical details of the structure
of the permutation of $2^N$ elements associated to the native Gray
code. In the last Section we add some final comments.

\section{The binary reflected Gray code}\label{Reflec}
The general definition of Gray code recalled in the Introduction
corresponds to what is known in the literature as a $(N,k)$-Gray
code and can in fact be fulfilled by very different orderings.
Consequently one can find in the literature a considerable variety
of members in the family of Gray codes and generalizations
\cite{varietat}. We are only interested in the length--$N$
\emph{binary} subclass which includes all $2^N$ words of $N$ bits
ordered in such a way that any two consecutive words differ in just
one bit. We will even restrict our attention to the binary reflected
Gray codes which are formally defined in a recursive way \cite{art}
:

Let $\Gamma_0$ denote the empty string and $\Gamma_N$ be the BRGC
sequence of $N$--bit strings, then
\begin{equation} \label{Knuth}
\Gamma_{N+1}=0\Gamma_N, 1\Gamma_{N}^{R}
\end{equation}
where $0\Gamma_N$ denotes the sequence $\Gamma_N$ with $0$ prefixed
to each string, and $1\Gamma_{N}^{R}$ denotes the sequence
$\Gamma_N$ in reverse order with $1$ prefixed to each string.

%\bigskip

\begin{table}[hb]
\begin{center}
\begin{tabular}{|c|r|r|r|c|}  \hline
           Decimal & Binary & Building proc.& Gray & $P_G$ \\ \hline\hline
           0 & 0000 & 0 & 0000 &            0 \\
           1 & 0001 & 1 & 0001 &            1 \\ \hline
           2 & 0010 & 11 & 0011 &            3 \\
           3 & 0011 & 10 & 0010 &            2 \\ \hline
           4 & 0100 & 110 & 0110 &            6 \\
           5 & 0101 & 111 & 0111 &            7 \\
           6 & 0110 & 101 & 0101 &            5 \\
           7 & 0111 & 100 & 0100 &            4 \\ \hline
           8 & 1000 & 1100 &  1100 &          12 \\
           9 & 1001 & 1101 &  1101 &          13 \\
          10 & 1010 & 1111 &  1111 &          15 \\
          11 & 1011 & 1110 &  1110 &          14 \\
          12 & 1100 & 1010 &  1010 &          10 \\
          13 & 1101 & 1011 &  1011 &          11 \\
          14 & 1110 & 1001 &  1001 &           9 \\
          15 & 1111 & 1000 &  1000 &           8 \\ \hline
\end{tabular}
\caption{Native Gray code construction for the first $2^4$ natural
numbers and its associated permutation $P_G$.} \label{TGray16}
\end{center}
\end{table}

Perhaps the easiest way to seize the idea that underlies the
reflected binary Gray code is through an example.
Table-\ref{TGray16} shows the first $16$ decimal integers
${0,1,\ldots, 15}$ and their corresponding binary Gray code with
$N=4$. The obvious first column lists the natural decimal ordering
we are familiar with. The second column, which we shall not use
explicitly in the following, is included just to explicitly
appreciate the differences between standard base-2 and Gray code
orderings. The latter is shown in the fourth column. A first look at
it reveals the aforementioned defining feature: adjacent entries
differ in one, and only one, bit. The third column is incorporated
to illustrate more clearly the building procedure from equation
(\ref{Knuth}). The fifth column, which is central to our discussion,
displays the decimal version of the previous column.

We have identified four blocks of rows in the Table-\ref{TGray16}
which will help in the explanation that follows, referred to the
third column. The first block with two rows can be considered the
seed for the Gray code, it represents $\Gamma_1$ from equation
(\ref{Knuth}). The next block includes the previous one together
with the added next two rows. When observing only the right-most bit
in these latter one sees simply the previous two rows in reverse
order, i.e. $\Gamma_{1}^{R}$. When $0$ and $1$ are added
respectively to the left we complete $\Gamma_2$. Observe that in
order not to obscure the illustration of the building process the
$0$ has not actually been written in the first two rows. The third
block adds $4$ new rows which accommodate the first $4$ rows in
reverse order and then the left-most $1$ is prefixed. The fourth
block illustrates the procedure again to obtain $\Gamma_4$ which is
given in full in the fourth column. The generalization to $2^N$
elements is straightforward. This is a recursive algorithmic way to
construct the Gray code table.

There are other possibilities for describing an algorithm leading to
Gray codes. For example, once we have chosen the number of bits,
$N$, we can start with an arbitrary string of length $N$. The next
term in the sequence is obtained by changing only one bit. From here
on the same procedure is repeated with the proviso that strings
which have already appeared in the listing are discarded. Observe
however that in this way we obtain certainly a binary Gray code,
although we cannot assure it is the native BRGC.

Apart from these recursive and constructive ways to obtain a BRGC of
given length there is also an explicit method \cite{Conder}. Let us
write the $k$th entry of the BRGC of length $N$ as the string
$b_1^N(k)b_2^N(k)\ldots b_N^N(k)$, then the $j$th bit (from left to
right) is given by the parity of a binomial coefficient by the
following expression which will be helpful in Section-\ref{permu} :
\begin{equation}\label{Conder}
    b_j^N(k)=^{2^N-2^{N-j}-1}C_{[2^N-2^{N-j-1}-(k+1)/2]} \qquad    ({\rm
    mod}\; 2)
\end{equation}
where $[\cdot]$ is the integer part function and $1\leq j\leq N$,
$0\leq k\leq 2^N-1$. A comment may be in order here. The previous
equation could give the wrong impression that the Gray code
representative of the natural number $k$ may sensibly change with
the total number of bits $N$. Either from equation (\ref{Conder}) or
directly from (\ref{Knuth}) one can easily see that this is not the
case. Increasing $N$ only adds $0$'s to the left of the
representative string.

The native code is cyclic in the sense that 16, in our example,
would go onto 0. More generally, the code is cyclic provided we deal
with a number of elements that is a power of $2$. It is worth
mentioning here that the exact number of cyclic Gray codes for
general $N$, which is related to the number of hamiltonian paths in
an hypercube, is still an open problem in combinatoric Analysis.
From the numeric stand point, the Gray code listings that we use
here were generated with the short Fortran routine in \cite{NR}.

Eventually, we are particularly interested  in interpreting the
entries in the Gray code column as natural numbers written in
base-2. Their decimal conversion is given in the last column of the
Table-\ref{TGray16}. Comparison of first and fifth columns admits
the following interpretation which we word in two obviously
equivalent ways:
\begin{enumerate}
\item The last column in Table-\ref{TGray16} is a specific
permutation of the $16$ elements $\{0,1,2,\ldots,15\}$. Or in
general, the length $N$ BRGC gives a permutation of the $2^N$
elements $\{0,1,2,\ldots,2^N-1\}$ which we denote by $P_G$.

\item The last column in Table-\ref{TGray16} defines a function of a discrete variable in
a definition domain with $16$ (in general $2^N$) points.
\end{enumerate}
The reason we have stated separately these, otherwise
interchangeable, sentences is because each of them suggests a
different line of approach. In fact, while the permutation
terminology is more amenable to combinatorial analysis, the
functional sense calls to mind the idea of discrete dynamical system
associated to the iteration of functions and related fractal
structures. Both interpretations will be considered in this paper
and yield interesting findings when considering $2^N$ elements with
$N\gg1$. By now, we stress the fact that all of our development
pivots about the correspondence between the first (decimal sequence
in standard order) and the fourth columns (decimal equivalent of the
binary Gray code entry) in Table-\ref{TGray16}. Gray codification
has been then just an intermediary in the process.

We introduce now some notation.  Let $\mathfrak{D}^{[N]}\equiv
\{D_n=n, n= 0,1,2,3,\ldots,2^N-1\}$ denote the sequence of the first
$2^N$ natural numbers in base-$10$ and $\mathfrak{G}^{[N]}\equiv \{
G_n, n=0,1,\ldots,2^N-1\}$, the decimal representation of the Gray
code, i.e., for $N=4$,  the last column in Table-\ref{TGray16}.
$P_G$ denotes the permutation that connects $\mathfrak{D}^{[N]}$ and
$\mathfrak{G}^{[N]}$ for a given $N$, i.e., $P_G$ is the element of
the symmetric group $S_{2^N}$ which acts on $\mathfrak{D}^{[N]}$
according to
\begin{equation}\
    P_G(n)=G_n,\quad n=0,1,\ldots,2^N-1.
\end{equation}
In the next two Sections we present a series of heuristic
considerations on $P_G$ deferring to Section \ref{permu} its more
technical details.

\section{A self-similar Gray set}\label{selfsim}
Let us now represent graphically the action of $P_G$. We consider
then the set $\mathfrak{A}_{N}$ of the $\mathbb{R}^2$ plane
consisting of the points with coordinates $(n,P_G(n))=(D_n,G_n)$,
$n=0,1,\ldots,2^N-1. $ In Figure-\ref{Gray-fig1}, we plot
$\mathfrak{G}^{[N]}$ \textsl{versus} $\mathfrak{D}^{[N]}$ up to
$N=18$, where we have connected successive points only as a visual
help. A direct view of the disconnected set is given in the lower
inset of the same figure. When no confusion may arise we shall
continue to refer to this set as the Gray code curve. We could
describe it, at first, as a kind of crinkly devil's staircase. In
view of the algorithmic way explained above to build up
$\mathfrak{G}^{[N]}$, it becomes clear the self-similar nature of
the plot: Increasing a new bit to the leftmost position of the Gray
number means to sum a certain power of two, and besides, the
preceding binary structure is inherited with a reversed ordering. In
the $\log_2-\log_2$ representation of the upper inset the scale
invariance is fully appreciated. In the present section we describe
the set $\mathfrak{A}_{N}$ for large $N$ from the point of view of
fractal geometry.

\subsection{The Gray code curve as a fractal}
\begin{figure}[th]
\begin{center}
\epsfxsize=4.6in {\epsfbox{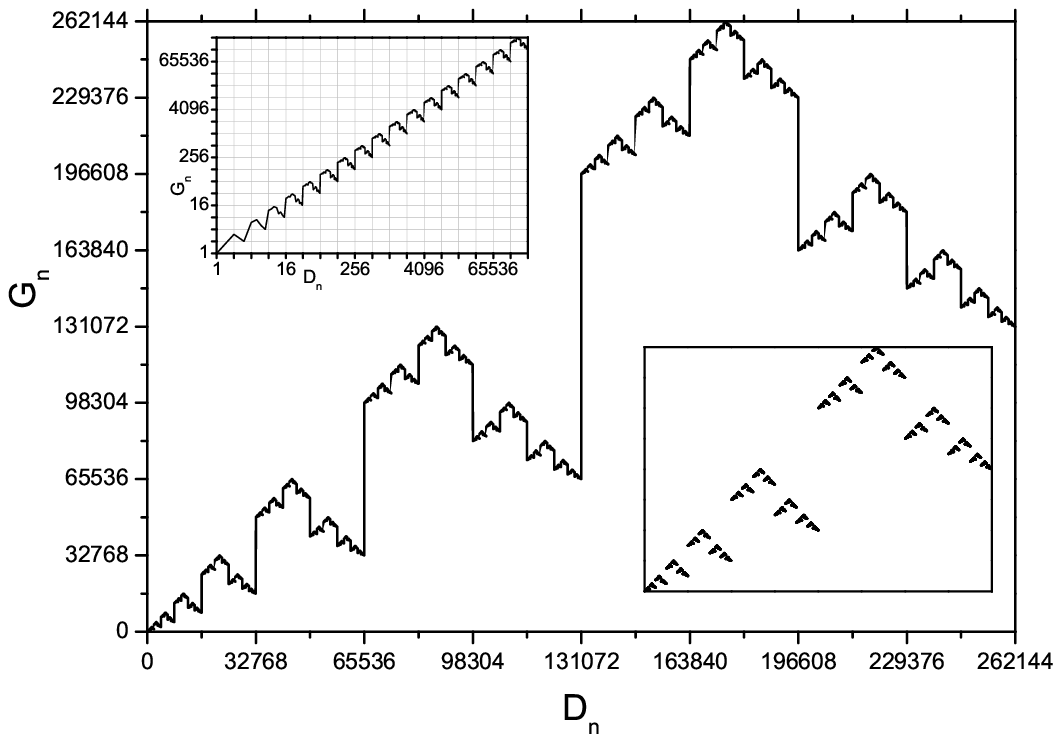}}
\end{center}
 \caption{ Self-similar Gray code curve}
 \label{Gray-fig1}

\bigskip

\begin{center}
\epsfxsize=4.6in {\epsfbox{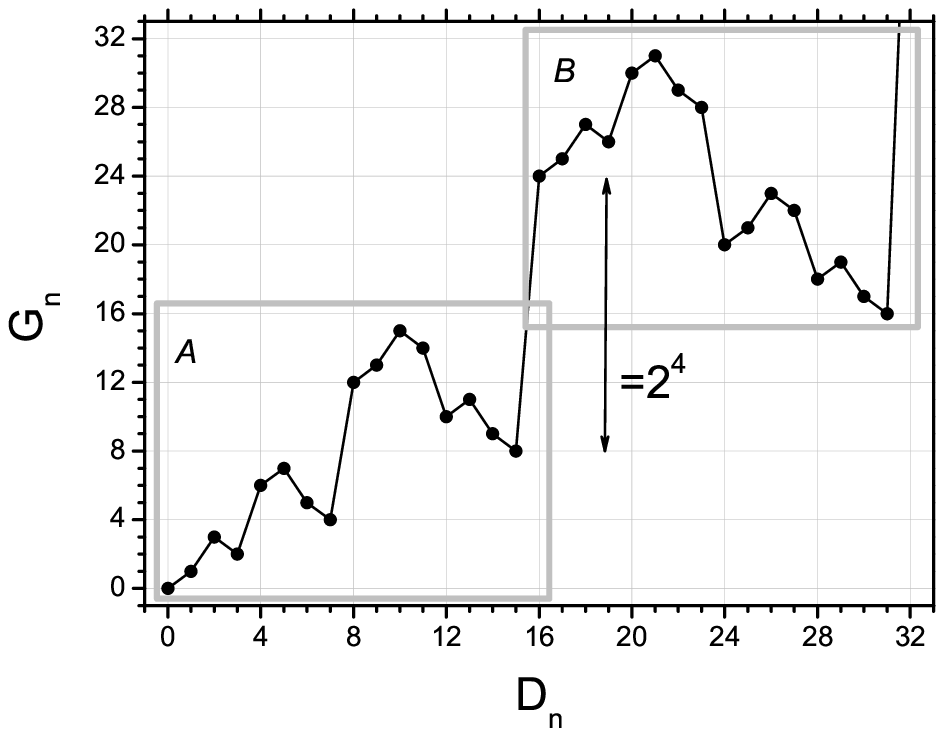}}
\end{center}
 \caption{ Self-similar construction of the Gray code curve }
 \label{Gray-IFS}

\end{figure}

The self-similarity of the Gray code curve may be made explicit.
Figure-\ref{Gray-IFS} contains its first $32$ points. Two subsets of
$16$ points each have been framed. The subset of points in frame B
is geometrically obtained from subset A by a procedure of the type
\textsl{replicate, reflect and translate}, as follows:
\begin{enumerate}
\item Make an adjacent copy of the set of points in panel A.
\item Reverse the abscissae order in the replicated copy, i.e.
substitute it by its image in a vertical mirror.
\item Add $2^{4}$ (= cardinal of the subset) units to every ordinate. This is panel
B.
\end{enumerate}
It is easy to see that inside plot frame A a similar procedure has
been followed starting with the eight first points. The seed of the
procedure was just   $\mathfrak{A}_0=\{(0,0)\}$.

To put this description in formulae let $\mathfrak{A}_k$ be
\begin{equation}\
    \mathfrak{A}_k=\{(n,P_G(n)) , n=0,1,\ldots,2^k-1\}\subset \
    \mathbb{R}^2,
\end{equation}
 the set of the first $2^k$ points of our curve, $k=1,2,\ldots$
Then
\begin{equation}
    \mathfrak{A}_{k+1}=\mathfrak{A}_k\cup F_k(\mathfrak{A}_k)
\end{equation}
where $F_k(\mathfrak{A}_k)$ is the image set of $\mathfrak{A}_k$
under the function
\begin{equation}\
    F_k\left(
          \begin{array}{c}
            x \\
            y \\
          \end{array}
        \right)=\left(
                  \begin{array}{cc}
                    -1 & 0 \\
                    0 & 1 \\
                  \end{array}
                \right)\left(
          \begin{array}{c}
            x \\
            y \\
          \end{array}
        \right)+\left(
          \begin{array}{c}
            2^{k+1}-1 \\
            2^k \\
          \end{array}
        \right)
\end{equation}
defined on $\mathbb{R}^2$ which corresponds to the previously
described manipulations. By this iterative procedure the whole curve
in Figure-\ref{Gray-fig1} grows out. From a purely geometric point
of view the previous function defines an affine transformation. But
its translation part depends on $k$, the generation index, a feature
to which we shall come back.

Alternatively we can inscribe the curve in the unit square with
vertices $(0,0),(1,0),(1,1),(0,1)$ by the following iterative
procedure. Let us denote by $\mathfrak{B}_0$ the set $\{(0,0)\}$,
incidentally $\mathfrak{B}_0=\mathfrak{A}_0$, and define
\begin{equation}\label {k_ifs}
    \mathfrak{B}_{k+1}=M(\mathfrak{B}_k)\cup L_k(\mathfrak{B}_k)
\end{equation}
where $M(\mathfrak{B}_k)$ and $L_k(\mathfrak{B}_k)$  are
respectively  the image sets of $\mathfrak{B}_k$ under the affine
(still $k$-dependent in the case of $L_k$) transformations
\begin{eqnarray}
   &M\left(
          \begin{array}{c}
            x \\
            y \\
          \end{array}
        \right)= \frac{1}{2}\left(
          \begin{array}{c}
            x \\
            y \\
          \end{array}
        \right)&  \\
   &L_k\left(
          \begin{array}{c}
            x \\
            y \\
          \end{array}
        \right)= \frac{1}{2}\left(
          \begin{array}{c}
            -x \\
            y \\
          \end{array}
        \right)+\frac{1}{2^k}\left(
          \begin{array}{c}
            {2^{k+1}-1} \\
            {2^k} \\
          \end{array}
        \right)&
\end{eqnarray}
which scale down, reflect and translate the curve. From all these
considerations the fractal flavor of the whole construction is
evident.

\subsection{IFS generation of the Gray curve} \label{IFS}
The procedures just described to construct the Gray set are clearly
reminiscent of the so called Iterated Functions Systems, very
popular in fractal building and image compression \cite{Barnsley}.
As it is well known a collection of functions are applied, either
following a deterministic plan or randomly, to a simple object. The
two more striking features of this approach are, on one hand, how
astonishingly complex images arise from very simple seeds. On the
other hand perfect coincidence in the final product of deterministic
and random processes, although more and more familiar every time
thanks in part to the popularization of chaos studies, is always
odd. Although not necessarily so, in most of the applications of IFS
the functions involved are a few affine transformations with
constant parameters.

It is natural then to look for an IFS generator of our set
$\mathfrak{A}_{N}$. The problem is, however, that one cannot find
constant parameters for a bidimensional affine transformation
capable of generating by iteration our fractal. The reason being the
$k$ dependence mentioned earlier. So we are led to look for more
general functions, i.e. non-linear ones,  of the coordinates, such
that when iterated produce the desired result. But the $k$
dependence continues to be a nuisance. The only way out is to
increase the dimension of the space we are working on from two to
three. This is exactly the same idea one takes advantage of in,
e.g., Classical Mechanics to convert a time dependent hamiltonian
system in an autonomous equivalent.

We then propose to extend $\mathbb{R}^2$ to $\mathbb{R}^3$ and take
as our IFS the following $\mathbb{R}^3 \rightarrow \mathbb{R}^3$
functions:
\begin{eqnarray}
              & f(x,y,z)=(\frac{x}{2},\frac{y}{2},z+1)&  \\
               &g(x,y,z)=(1-\frac{x}{2}-2^{-z},\frac{1+y}{2},z+1).&
\end{eqnarray}

The price one has to pay for getting rid of the $k$ dependence is
the higher dimension we work on and the non-linear character of the
function $g$ involved. Needless to say that for representation
purposes we only plot the projection on the $xy$ plane.  Anyway,
equation (\ref {k_ifs}) changes now to
\begin{equation}
    \mathfrak{B}_{k+1}=f(\mathfrak{B}_k)\cup g(\mathfrak{B}_k).
\end{equation}
Observe that the action of $f$ and $g$ on the $z$ coordinate
prevents the existence of fixed points when each of these functions
is iterated on its own. In the $xy$ plane when $f$ is repeatedly
iterated the results tend rapidly to the origin. It is only a little
bit more complicated to see that iteration of $g$ leads in the limit
to the point $(1,1)$. The combined and competing iteration of the
two functions is what produces the final figure.
 The set \{f,g\} constitutes a
\emph{bona fide} IFS. The system is deterministic if $f$ and $g$ are
repeatedly applied one after the other. When this is done starting
with $(0,0)$, then the whole set $\mathfrak{A}_{N}$ is generated. It
is a random IFS if $f$ and $g$ are iterated with prescribed
probability $\frac{1}{2}$ on an arbitrary initial point in the unit
square. $\mathfrak{A}_{N}$ is obtained as the final attractor of the
process.

\section{A time series from BRGC}\label{timeseries}

Our proposal now is to interpret $G_n$ as successive values in a
time series. Conventional statistics (average values, standard
deviations and other indicators calculated from higher moments) are
of no particular interest for us here because they depend solely of
the numerical values $G_n$ and these are simply the first $2^N$
natural numbers. The order in which the different values appear does
not play any role in the computation of those statistics. The
information from the time series obtained in this way could be
termed as static. Instead, we are interested in the dynamics
represented by the series. The ordering, then, matters and we need
other indicators. With this aim we analyze in the present Section
the series of increments $\{ \Delta_n\}$ with
$\Delta_n=G_n-G_{n-1}$, $n=1,2,\ldots$

\subsection{Statistics on $\Delta_n$} \label{Stat-D}
\begin{figure}[th]
\begin{center}
\epsfxsize=4.6in {\epsfbox{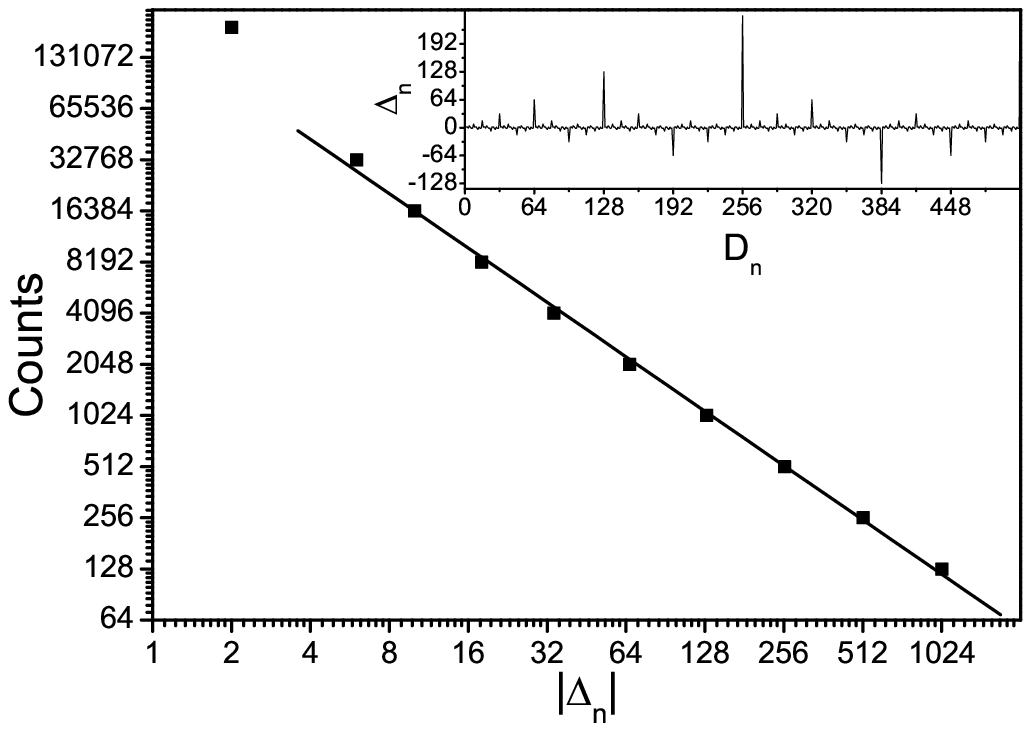}}
\end{center}
 \caption{ Histogram of the increments $\{ |\Delta_n|\}$ in $\log_2-\log_2$ scale}
 \label{Gray-powerlaw}

\bigskip

\begin{center}
\epsfxsize=4.6in {\epsfbox{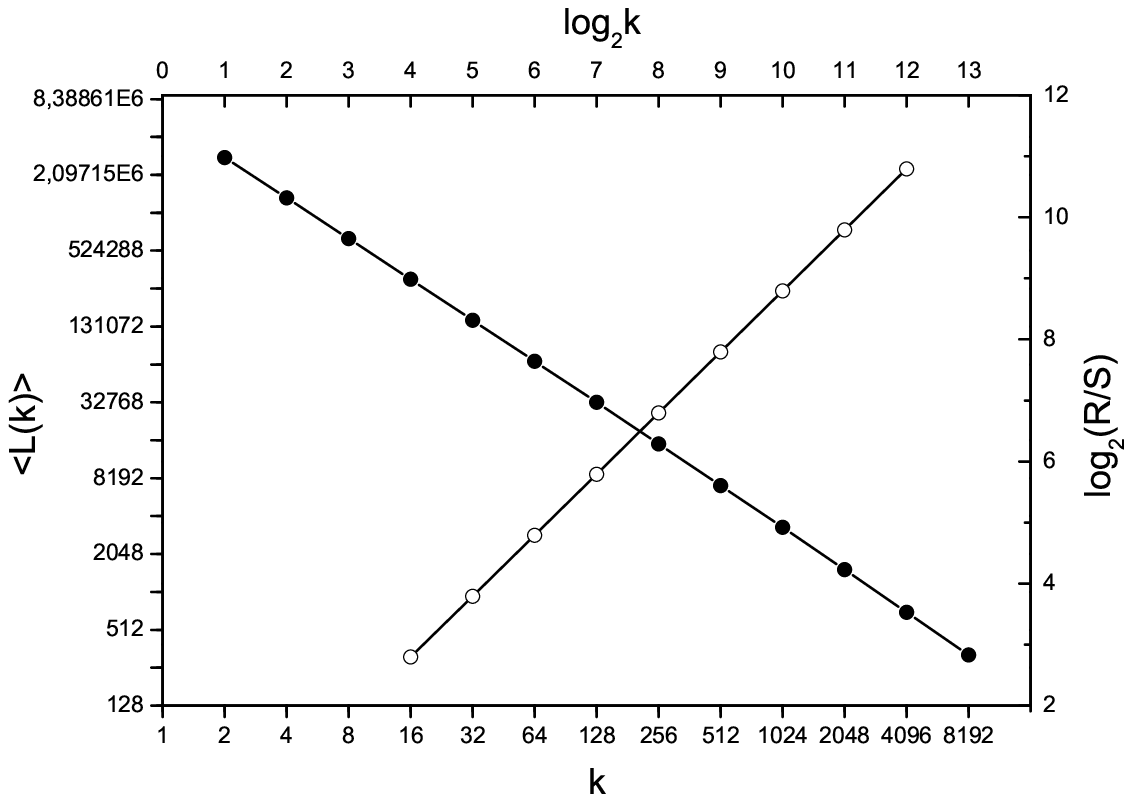}}
\end{center}
 \caption{ Determination of the fractal dimension (left scale, solid dots)
 and Hurst exponent (right scale, circles)}
 \label{Gray-FDH}
\end{figure}

The inset in Figure-\ref{Gray-powerlaw} shows a sample of the series
of increments $\{ \Delta_n\}$. By its very definition the average of
the first $m$ values is $G_m/m$ and because it is always positive
the Gray curve is globally increasing. The statistics on the set of
values $\{ |\Delta_n|\}$ reveals that their frequency obeys a power
law, with the exception of the very small values (see main graph in
Figure-\ref{Gray-powerlaw}). As a matter of fact, the frequency
histogram goes like $\sim 1/|\Delta|$, which means that the
probability of encountering a jump $|\Delta |< \delta$ grows as
$\log \delta$. One could express this fact by saying that the values
$G_n$ are far from equidistant. By contrast, it seems appropriate to
mention here that in its binary form, BRGC is characterized by any
two consecutive elements being equally spaced by a unit Hamming
distance.
\subsection{Residence times from $\Delta_n$}
An often encountered statistic when analyzing time series is the
so-called residence  time, or sojourn time, $\tau$. To deal with it,
one has to define a dichotomic property of the system such as
crossing a predefined threshold and keep track of the time elapsed
between successive crossings. In our case we focus our attention on
changes in sign of $\Delta_n$. Accordingly we define the residence
times as the length of the sequences in which $\Delta_n$ keeps its
sign. If it stays just one unit in the, say, positive region, we
define its sojourn time to be just one unit.

When we collect the statistic for residence times in the set of
values of $\Delta_n$ the fact which first calls our attention is
that $\tau$ can only take three different values: $1, 2$ or $3$.
Suppose $\Delta_n$ has changed sign with respect to $\Delta_{n-1}$,
then there are only three possibilities: either $\Delta_{n+1}$ has
sign opposite to that of $\Delta_n$ (then $\tau=1$), or
$\Delta_{n+1}$ and $\Delta_n$ have equal sign but different from
that of $\Delta_{n+2}$ (which corresponds to $\tau=2$), or
$\Delta_n$, $\Delta_{n+1}$ and $\Delta_{n+2}$ have the same sign
which necessarily changes in $\Delta_{n+3}$ (giving $\tau=3$). Never
more than two consecutive steps can be given without encountering a
change of sign. As a matter of fact the statistics reveals more than
that: the probabilities (relative frequencies of appearance) for the
values of $\tau$ are exactly
\begin{equation}
     p(1)=0.25, p(2)=0.5, p(3)=0.25, p(\tau >3)=0.
\end{equation}
This behavior of the residence times contrasts with other known
situations. For example, exponentially decreasing probability
distributions are common in chaotic systems.

It is worth mentioning that the statistics of residence times we are
treating provides information only about changes of sign in
$\Delta_n$. It says nothing about the direction (upwards or
downwards) of the change. The fact that, as mentioned earlier, the
average of increments $\{ \Delta_n\}$ is positive suggests the
increasing overall character of our curve. By simply looking at  it
at any scale one corroborates that it is globally increasing, in
spite of its ups and downs we perceive in more detailed graphs. In
the jargon of time series analysis one says that this reflects
extremely persistent dynamics which we analyze next from the Hurst
exponent point of view.

\subsection{Hurst's $R/S$ analysis of $\Delta_n$}
The so-called re-scaled range statistical analysis, $R/S$ analysis,
is a method designed to reveal long-run correlations or
anti-correlations in a complex process. The idea was first
introduced in the analysis of some hydrologic time series, but has
been also applied in other studies in climatology, geophysics,
econometrics,\ldots

Following Feder \cite{Feder}, we define the average increment over a
period of $k$ units as
\begin{equation}
\langle \Delta \rangle _k=\frac{1}{k}\sum_{r=1}^k \Delta_r= G_k/k
\end{equation}
The local accumulated departure of $\Delta_n$ from the mean $\langle
\Delta \rangle _k$ is then
\begin{equation}
X(n,k)\equiv \sum_{r=1}^n [\Delta_r-\langle \Delta \rangle
_k]=G_n-\frac{n}{k}G_k
\end{equation}
The range of $X(n,k)$ over this period reads
\begin{equation}
R(k)\equiv \max X(n,k)- \min X(n,k)
\end{equation}
where the extremal functions are evaluated in the interval $1\le
n\le k$. The range increases with $k$. When properly normalized to
the standard deviation
\begin{equation}
S=\left(\frac{1}{n}\sum_{r=1}^n [\Delta_r-\langle \Delta \rangle
_k]^2\right)^{1/2}
\end{equation}
then Hurst found that the re-scaled range $R/S$ is well described by
the power law
\begin{equation}
R/S=(k/2)^H
\end{equation}
where $H$ is called  \emph{Hurst exponent}.  Purely random series
give $H=1/2$. Curves with $H>1/2$ ($H<1/2$) witness a long run
correlation of persistent (anti-persistent) character. This feature
means that positive changes are more likely followed by positive
(negative) ones, whereas for random phenomena both results for the
next move are always equiprobable. Hurst \cite{Hurst} found that
most natural phenomena give rise to $H\simeq 0.7$, which is
sometimes referred to as Hurst phenomenon.

In the case of the Gray curve a Hurst--like analysis provides a
value $H=0.97(3)$, revealing the extreme persistence of the series
of increments, as pointed out above. This analysis appears in
Figure-\ref{Gray-FDH} where $R/S$ (straight line with open circles)
is to be read on the right scale.

\subsection{Fractal dimension of the Gray curve}
We follow a technique by Higuchi \cite{Higuchi} to measure the
fractal dimension of a set of points $\{n,G_n\}$,
$n=0,1,2,\ldots,2^N$, forming the graph of a function like that in
Figure-\ref{Gray-fig1}. The idea consists in measuring the average
length $\langle L(k) \rangle $ of samples with $k$ points each.
Then, if $\langle L(k) \rangle \varpropto k^{-d}$, the curve is
fractal with dimension $d$. For instance, a time series obtained
with pure noise should give $d=1.5$, whereas for a regular curve
$d=1$. The method recalls the so-called Richardson plot in fractal
geometry.

In Figure-\ref{Gray-FDH} we have plotted in doubly logarithmic scale
$\langle L(k) \rangle $ against $k$, with $N=18$ (straight line with
solid dots). The straight line is fitted to the points by the least
square method and gives a slope $d=1.09(1)$. Here the error estimate
comes from the stability of the result with respect to the number of
data rather than from the fitting procedure.

These values for $H$ and $d$ are consistent with the connection
formula relating $H$ and $d$ which reads: $d=2-H$.

\subsection{Return plot of the Gray curve}
\begin{figure}[th]
\begin{center}
\epsfxsize=4.6in {\epsfbox{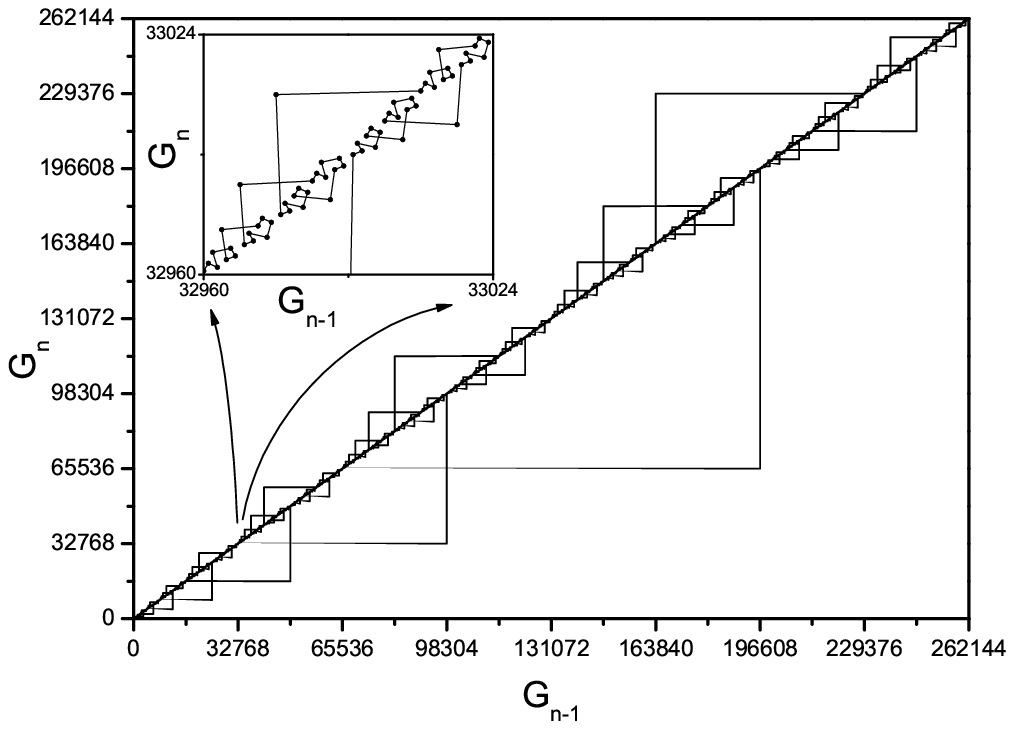}}
\end{center}
 \caption{ 2D trajectory for the Gray curve and zoom of the marked area }
 \label{Gray-fig3}

\vspace{2cm}

\begin{center}
\epsfxsize=4.6in {\epsfbox{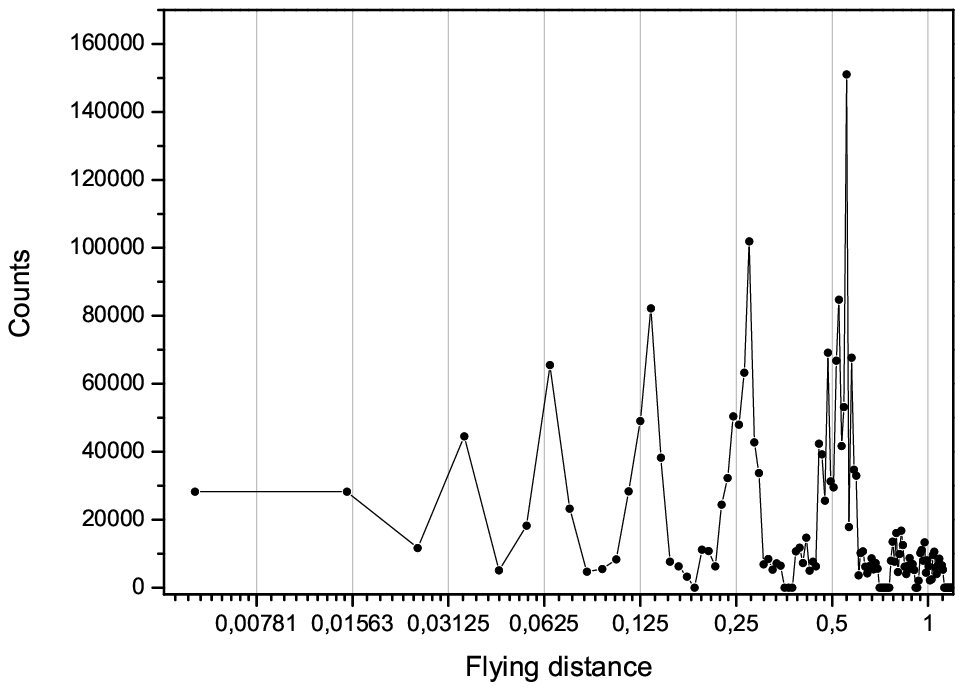}}
\end{center}
 \caption{ Histogram of {\sl Flying Distances in the random IFS Gray fractal
 with $N=22$ and bin size $0.02$ }}
 \label{Fig-Fly}

\end{figure}

A common tool in nonlinear analysis of time series consists in
giving a 2D- representation of the 1D time series called return
plot. A return plot, or first--return map, of the sequence $\{G_n\}$
is a 2D--plot of the points $\{(G_{n-1}, G_{n})\}$. Thus, for
instance, a true random sequence gives rise to a uniformly scattered
plot. Chaotic sequences where the generation of the $n$th term
involves only the $(n-1)$th term produce 1D--patterns. In the
present case, the return plot for the Gray curve is in
Figure-\ref{Gray-fig3}. We have drawn not only the points but also a
path connecting them in order to realize the variety of jumps
involved, in accordance with the set of values $\{\Delta_n\}$. The
inset in Figure-\ref{Gray-fig3} is a zoom of the indicated zone
showing the detail of the curly structure of the curve. It is clear
how strongly the natural numbers sequence gets intermingled in the
permutation $P_G$: The recurrence plot for the set $\{ D_n\}$ is
simply a straight line like $y=x+1$.

\subsection{Flying distances in the random IFS Gray fractal}
At the end of subsection \ref{IFS} it was pointed out that the Gray
fractal can be obtained as the attractor of a random process of
iteration of functions $f$ and $g$. Here we investigate the
statistics obeyed by the distances traveled in every move. This
point of view quits the static perspective on the fractal and enters
a new kind of dynamics.

Starting from an arbitrary point in the unit square,  runs of
$2^{22}$ points were carried out. We have retained just the last
$2^{21}$ points, and the previous ones were considered as a
transient. The idea is to lose any memory about the initial position
by getting close enough to the true attractor. A histogram of the
Euclidean distances traveled  between two consecutive points in a
random walk on the Gray fractal is shown in Figure-\ref{Fig-Fly}. It
is noteworthy the presence of equally spaced peaks on the log-scale.
The same is true for the bottom of valleys. This phenomenon means
that the geometry of the fractal selects (or avoids) some particular
flying distances along its random generation. The specific mechanism
underlying this selection rule is unclear, as according to the
statistic of $\Delta_n$ in subsection-\ref{Stat-D} all length
distances are present in the fractal itself.

\section{The native Gray permutation}\label{permu}
\begin{figure}
\begin{center}
%\centering
   \epsfxsize=4.1in {\epsfbox{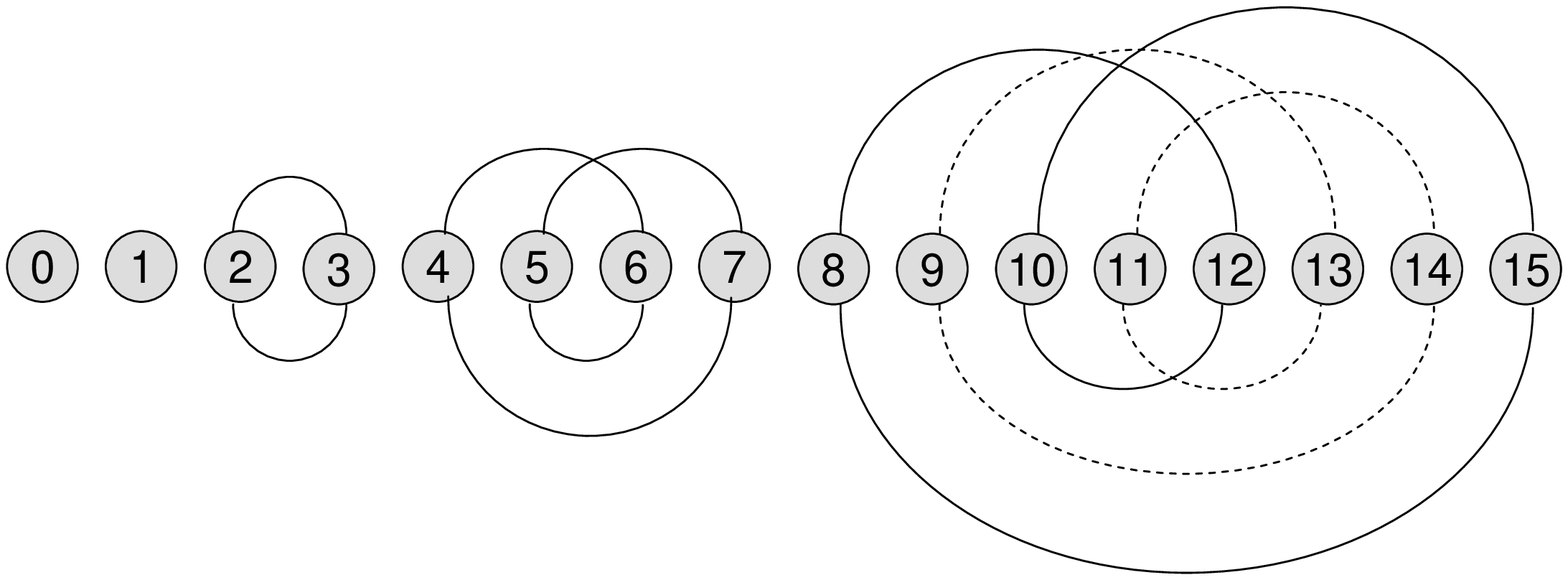}}
   \epsfxsize=4.1in {\epsfbox{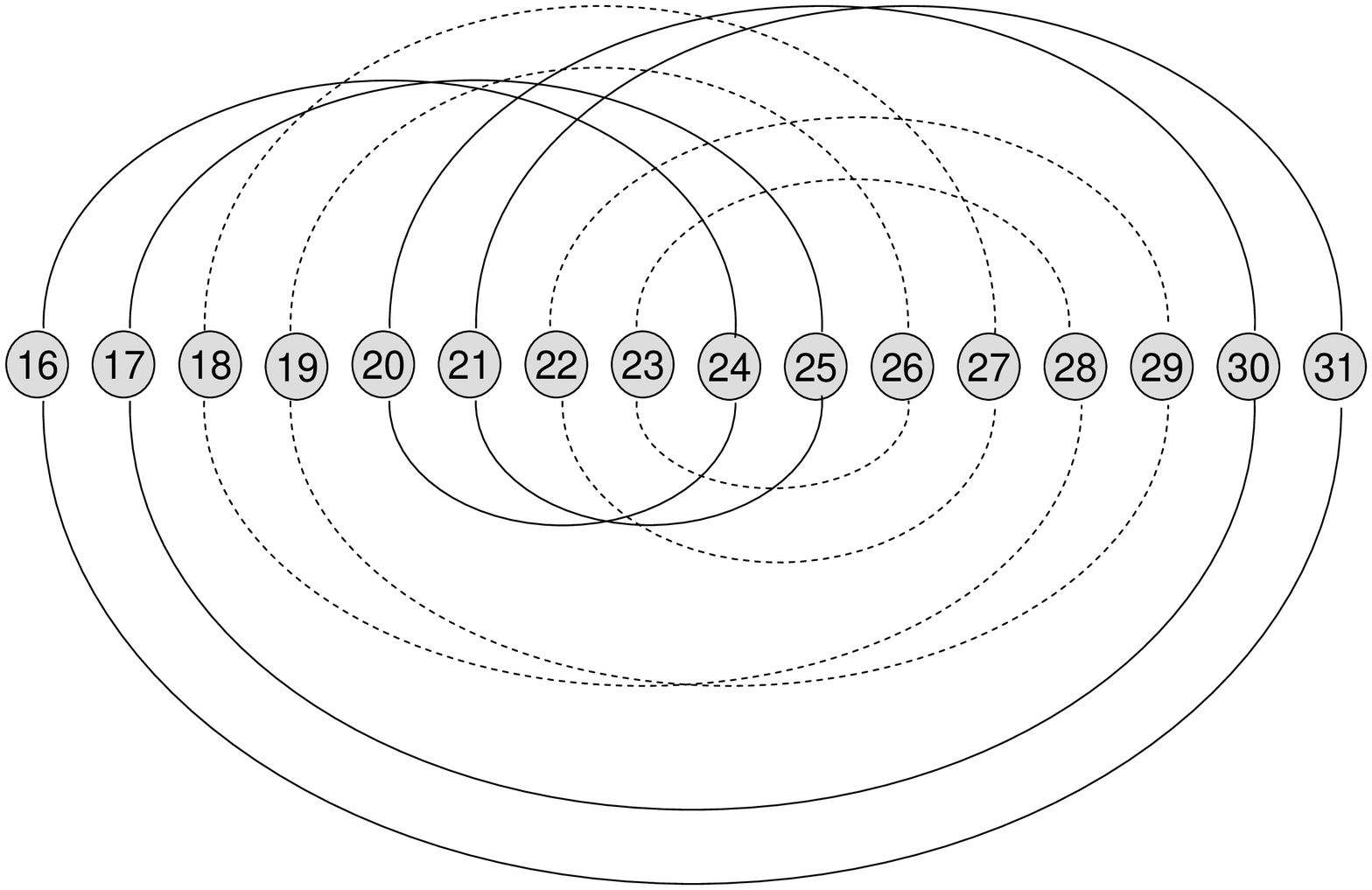}}
\end{center}
 \caption{Orbits for Gray permutation of elements $0$ to $31$}
 \label{orbits}

\vspace{2cm}

\begin{center}
\epsfxsize=4.6in {\epsfbox{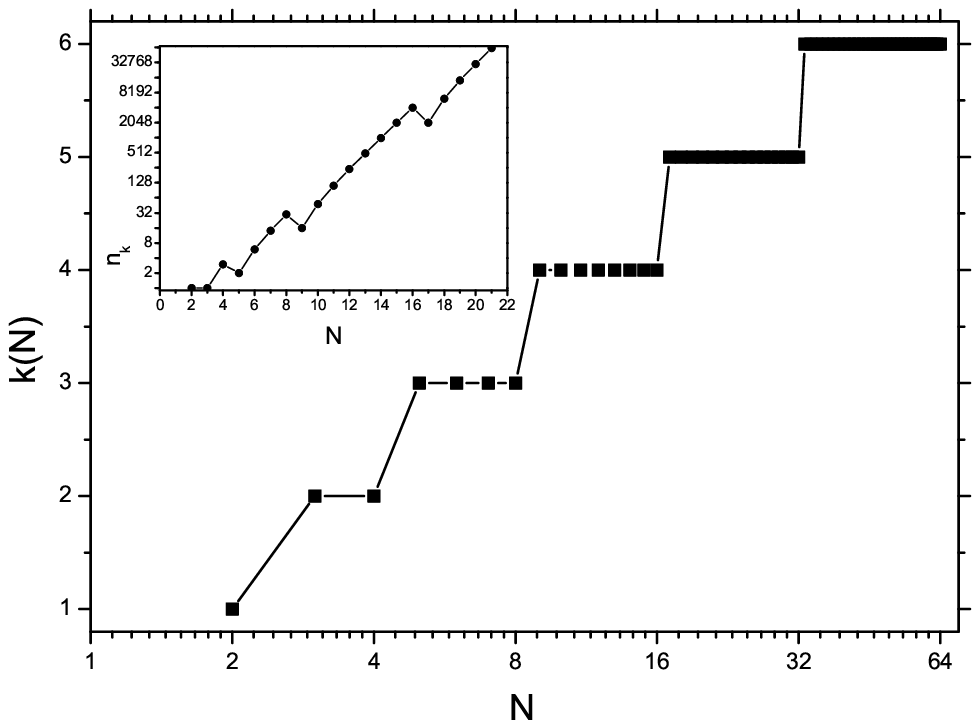}}
\end{center}
 \caption{ Statistic of cycles as a function of $N$}
 \label{Fig-NK}

\end{figure}

In this section we collect some properties of the permutation $P_G$
which have been instrumental in most of the considerations presented
so far. In correspondence with the name \emph{native Gray code}
sometimes used for the BRGC we call $P_G$ the \emph{native Gray
permutation}. A little experimentation and some reflection lead to
the following observations:
\begin{enumerate}
\item The recurrent definition of the permutation $P_{G}$ as a function with
domain $\mathfrak{D}^{[N]}$ is
\begin{equation}
P_{G}(2^{n}+m) =2^{n}+P_{G}(2^{n}-m-1)
\end{equation}
with $P_{G}(0)   =0$, and where $ n=0,1,\ldots, N-1$,
$m=0,1,\ldots,2^{n}-1$. Particularly interesting values are
$P_G(2^n-1)=2^{n-1}$ and $P_G(2^n)=3 \cdot 2^{n-1}, n\geq1$.

\item An explicit expression for $P_G(n)$ can be deduced from
equation (\ref{Conder}). Given $n\in \mathbb{N}$, $n\geq1$, the
minimum number of bits necessary to write it in base-2 is $v(n)$
given by
\begin{equation*}\nonumber
     v(n)=[\log_2(n)]+1.
\end{equation*}
Then we have
\begin{equation}\label{permuta}
    P_G(n)=\sum_{j=1}^{v(n)}b_j^{v(n)}(n) \cdot 2^{v(n)-j}
\end{equation}
where $b_j^{v(n)}(n)$ has been defined in Section-\ref{Reflec}

\item The equation $P_{G}(x)=x$ has, for any value of $N$, only two solutions:
$x=0$ and $x=1.$ It is to say, $0$ and $1$ are the only elements of
$\mathfrak{D}^{[N]}$ which $P_{G}$ leaves invariant. All other
elements of $\mathfrak{D}^{[N]}$ under the action of $P_{G}$ are
grouped in cycles, but only cycles of length $2^{l}$ will appear.
In symbols that means that we can factor out $P_{G}$ in cycles in the form%
\begin{equation}
P_{G}=\prod\limits_{l=0}^{k(N)}c_{l}^{n_{l}}
\end{equation}
where $c_{l}$ represents a cycle of $2^{l}$ elements of which
$n_{l}$ different specimens appear in $P_{G}$. $k(N)$ indicates that
the length of the longest cycle or cycles is $2^{k(N)}$.

\item The distribution of the $2^{N}$ elements of $\mathfrak{D}^{[N]}$ in cycles follows a
strict procedure: they fill cycles $c_{l}$ in increasing order of
$l$. No $2^{l}-$cycle appears until all allowable $2^{l-1}-$cycles
are filled up. Moreover, $n_{l}$ has a maximum value independent of
$N.$ In other words, once a certain number of $2^{l}-$cycles has
been reached no matter how big $N$ becomes no other $2^{l}-$cycle
will be formed. Furthermore, when $N$ increases the factorization of
$P_{G}$ keeps the structure corresponding to the lower values of
$N$. The new cycles formed are as long as the longest ones for
previous values of $N$ or longer. This property confers to $P_G$ a
certain type of shell structure.

\item $2^{l}-$cycles will appear for the first time for
$N=2^{l-1}+1.$ In the language of the theory of dynamical systems
one could say that at these values of $N$ a period doubling
bifurcation appears. This way of speaking has to be considered
simply as an analogy and cannot be taken too literally. It is
illustrated with an example in Figure-\ref{orbits} which is
self-explanatory.
\end{enumerate}

With all these considerations in mind a careful book-keeping shows
that the length of the longest cycle in $P_G$ is $2^{k(N)}$ with
\begin{equation}
k(N)=\left[ \frac{\ln(N-1)}{\ln2}\right] +1=[\log_{2}(N-1)]+1,\qquad
N\geq2
\end{equation}
where $[\cdot]$ stands for integer part. Obviously for $N=1$ we have
$k(1)=0$ and, for any $N,$ $n_{0}=2.$

According to observation (iv) above, the maximum value of $n_{l}$,
$n_{l}^{max}$ is, for $l<k(N)$,
independent of $N$ and it is not difficult to see that%
\begin{equation}
n_{l}^{max}=2^{(2^{l-1}-l)}(2^{(2^{l-1})}-1),\qquad0<l<k(N)
\end{equation}
while the number of longest cycles is
\begin{eqnarray}
n_{k}&=&2^{(2^{k-1}-k)}(2^{p}-1) \nonumber \\
p &=&N-2^{k-1}
\end{eqnarray}
where in these formulae $k(N)$ has been for simplicity written as
$k$. Observe that in this last case the dependence on $N$ is
explicit. As a test one can check that
\begin{equation}
\sum_{l=0}^{k(N)}2^{l}n_{l}=2^{N}
\end{equation}

To fully appreciate the complexity of the cycle structure of the
Gray permutation $P_{G}$ we show in Table-\ref{Tciclos} a few cases
of values characterizing that structure. It gives the complete cycle
structure of $P_{G}$ up to $N=10$ and includes also the total number
of cycles, $N_{c}$, and the average cycle length, $l_{av}$.

\begin{table}[h]
\begin{center}
\begin{tabular}
[c]{|c|c|c|c|c|c|c|c|c|}\hline $N$ & $2^N$ & $n_{0}$ & $n_{1}$ &
$n_{2}$ & $n_{3}$ & $n_{4}$ & $N_{c}$ & $l_{av}$\\\hline $1$ & $2$
&2 & $0$ & $0$ & $0$ & $0$ & $2$ & $1$\\
$2$ & $4$ & $2$ & $1$ & $0$ & $0$ & $0$ & $3$ & $1.33$\\
$3$ & $8$ & $2$ & $1$ & $1$ & $0$ & $0$ & $4$ & $2$\\
$4$ & $16$ & $2$ & $1$ & $3$ & $0$ & $0$ & $6$ & $2.66$\\
$5$ & $32$ & $2$ & $1$ & $3$ & $2$ & $0$ & $8$ & $4$\\
$6$ & $64$ & $2$ & $1$ & $3$ & $6$ & $0$ & $12$ & $5.33$\\
$7$ & $128$ & $2$ & $1$ & $3$ & $14$ & $0$ & $20$ & $6.4$\\
$8$ & $256$ & $2$ & $1$ & $3$ & $30$ & $0$ & $36$ & $7.11$\\
$9$ & $512$ & $2$ & $1$ & $3$ & $30$ & $16$ & $52$ & $9.85$\\
$10$ & $1024$ & $2$ & $1$ & $3$ & $30$ & $48$ & $84$ &
$12.19$\\\hline
\end{tabular}
\end{center}
\caption{Cycle structure of permutation $P_G$} \label{Tciclos}
\end{table}

The reminiscence of the atomic shell-model mentioned earlier is of
course purely symbolic but it is reinforced when observing
Figure-\ref{Fig-NK}. In it we represent $k(N)$ as a function of $N$
and one can see that the jumps occur for values of $N$ which are
exact powers of $2$. The inset shows that precisely at these values
the number of longest cycles falls down.

Of course the general formulae we have deduced do not give
explicitly the elements in each cycle like for example, for $N=4$,
\[
P_{G}=(0)(1)(2,3)(4,6,5,7)(8,12,10,15)(9,13,11,14),
\]
but this information is much less relevant for our purposes.

It is interesting to notice the somewhat distinguished role of $P_G$
in the symmetric group $S_{2^N}$. It is clear that from the
information given in this section one can easily draw the Young
diagram corresponding to the equivalence class containing $P_G$ in
the group $S_{2^N}$. It is also obvious that  binary Gray codes
other than BRGC originate also permutations in $S_{2^N}$, but these
may have a cycle structure very different to that of $P_G$ and
consequently behave in other ways. On the other side any element
$Q\in S_{2^N}$ generates trivially a binary code, but in general it
is not of the Gray type.

\section{Final comments}\label{conclu}
Among combinatorics problems with physical applications the study of
permutations has always been of particular interest. As recent
exponents of it, in \cite{randomwalk} and \cite{randomwalk1} random
permutations are used in connection with stochastic processes. On
the other hand Gray codes have also met applications in different
branches of Physics, as widely shown in the Introduction.

From the point of view of combinatorics or graph theory
relationships between this type of codes and permutations are
manifold and are being explored in different directions in
mathematical literature \cite{mate2,mate3,mate1}. But from physics
standpoint no such kinship seems to have been studied in detail. In
this paper we have analyzed the permutation associated with the
binary reflected Gray code from the perspective, and with the
techniques, of a dynamical system and its corresponding time series.
In so doing we have been led in a natural way to enhance some
fractal aspects of the system. One may hope that this line of
approach could be helpful in understanding a more dynamical role of
permutations.

The results of the analysis carried out in this paper involve only
integer arithmetic, some book-keeping and graphical considerations
and so constitute, in our opinion, appropriate material for
pedagogical purposes.

As a final summary we would like to mention as a  particularly
appealing result of our approach the generation of the permutation
$P_G$ by an iterated function system applied either in a
deterministic or a random way. Also worth mentioning is the
discovery of strong selection rules acting on the system which
manifest themselves in the distribution of residence times with only
three allowed values, and of flying distances. Which, if any, of
these results apply to other permutations associated with other
codes may perhaps be worth considering in the future. As would be a
deeper understanding of the selection rules mentioned before.

\ack  This work has been partially supported by contracts
MCyT-BMF2001-0262 and GV2003-002, Spain.

%\newpage

\newpage

\newpage

\newpage

\end{document}